\documentclass[aps,
reprint]{revtex4-2}

\usepackage{graphicx}
\usepackage{float}

\usepackage[utf8]{inputenc}
\usepackage{comment}
\usepackage{dcolumn}
\usepackage{bm}

\usepackage{xcolor}
\usepackage{comment}
\usepackage{soul}

\usepackage{hyperref}
\usepackage{adjustbox}
\hypersetup{
    colorlinks=true,
    linkcolor=blue,
    filecolor=magenta,      
    urlcolor=cyan,
    pdftitle={Overleaf Example},
    pdfpagemode=FullScreen,
    }

\usepackage{natbib}
\usepackage{amsmath,amssymb}

\begin{document}

\title{Resource overheads and attainable rates for trapped-ion lattice surgery}
\author{Hudson Leone}
\affiliation{
     Centre for Quantum Software and Information -- University of Technology Sydney
     }
     \affiliation{Centre for quantum computation and Communication Technology (CQC2T)}
\author{Thinh Le}
\affiliation{
     Centre for Quantum Software and Information -- University of Technology Sydney
     }
\author{S.\ Srikara}
\affiliation{
     Centre for Quantum Software and Information -- University of Technology Sydney
     }
\author{Simon Devitt}
 \email{Simon.Devitt@uts.edu.au}
     \affiliation{
     Centre for Quantum Software and Information -- University of Technology Sydney
     }
     \affiliation{InstituteQ, Aalto University, 02150 Espoo, Finland.}

\date{\today}

\begin{abstract}
We present estimates for the number of ions needed to implement fault-tolerant lattice surgery between spatially separated trapped-ion surface codes.
Additionally, we determine attainable lattice surgery rates given a number of dedicated ``communication ions" per logical qubit.
Because our analysis depends heavily on the rate that syndrome extraction cycles take place, we survey the state-of the art and propose three possible cycle times between $10$ and $1000 \mu s$ that we could reasonably see realised provided certain technological milestones are met.
Consequently, our numerical results indicate that hundreds of resource ions will be needed for lattice surgery in the slowest case, while close to a hundred thousand will be needed in the fastest case.
The main factor contributing to these prohibitive estimates is the limited rate that ions can be coupled across traps.
Our results indicate an urgent need for optical coupling to improve by one or more orders of magnitude for trapped-ion quantum computers to scale.

\end{abstract}

\maketitle

\section{Introduction}

Trapped ions are among the best studied and most technologically mature type of qubits to date;
Their long coherence times and high-fidelity gates alone justify them as a candidate qubit for scalable quantum computing \cite{review}.
How a quantum computer is scaled will depend on its underlying architecture.
Broadly speaking, an architecture may be \textit{monolithic} \cite{Lekitsch_17} or \textit{modular} \cite{Nemoto_14}.
A monolithic architecture is scaled by increasing the size of the chip, while a modular architecture is scaled by increasing the number of chips.
Physical constraints and routing overheads generally cap the number of qubits that a monolithic architecture can reasonably support, which makes modularity something of an informal requirement when scaling quantum computers.
Another requirement for scalability is error correction \cite{Roffe2019}.
Industrial applications for quantum computing require programs to run for hours or even days.
Since quantum operations introduce small amounts of error into the ion states, the computational qubits must therefore be encoded and periodically corrected for lengthy computations to succeed.

Based on these considerations, we expect that scalable quantum computers will be both \textit{modular} and \textit{error corrected}.
One obvious disadvantage of the modular architecture is that two-qubit operations are \textit{not intrinsically possible} between qubits that live in separate modules.
Instead, some amount of entanglement has to be shared between modules as a resource for \textit{state} or \textit{gate teleportation} \cite{Chou2018}.
This process is complicated by the fact that entanglement distribution is \textit{probabilistic} and \textit{noisy}.
To say that distribution is probabilistic means there is a significant chance of failure when attempting to entangle two physical qubits.
For inter-modular two qubit operations to be reliable, this means that a large number of \textit{communication qubits} will be required as an overhead to ensure that enough entanglement is collected.
On the other hand, to say that distribution is noisy means that errors are inadvertently introduced into the entangled states which must be corrected before any teleportation can take place.
This correction is done through \textit{entanglement purification} \cite{Bennett96} which non-deterministically reduces a large ensemble of weakly entangled pairs into a smaller ensemble of strongly entangled pairs.

\begin{figure}
    \includegraphics[scale=0.20]{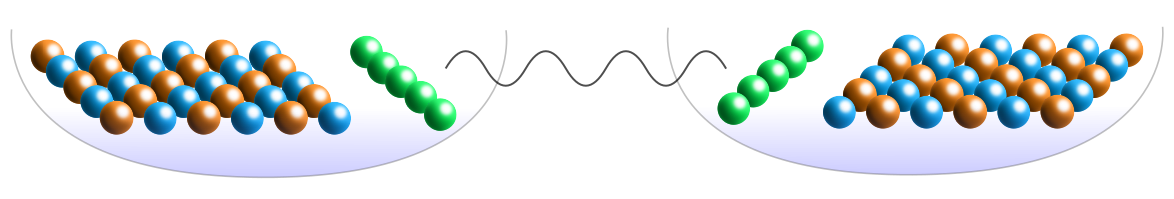}
    \caption{\label{fig:TrappedIonDrawing} 
    A collection of trapped ions in two separate Elementary Logical Units (ELUs). 
    Some of the ions in each trap are used to encode a logical surface code qubit while other \textit{communication ions} collect and refine entanglement to be used for a two-qubit \textit{lattice surgery} operation.
    }
\end{figure}

In this paper, we conduct resource analysis on the number of ions needed to implement \textit{reliable} two qubit operations between logically encoded qubits in different modules at different speeds.
Specifically, we limit our attention to the \textit{lattice-surgery} operation between \textit{surface-code} encoded qubits (See Fig.\ \ref{fig:TrappedIonDrawing} for a simplified schematic).
This analysis is significantly influenced by the speed at which lattice surgery can be performed.
Faster operations are desirable of course, but myriad factors limit the attainable rate.
To ground our analysis, we survey the state-of-the-art in trapped-ion technologies and synthesise this information to propose \textit{three possible surgery times} together with a rubric of technological milestones that are necessary to achieve each one.
Additionally, we determine the attainable rates for lattice surgery operations given a fixed number of ions.

\section{Background}

\subsection{Surface code and lattice surgery}

An $[[n,k,d]]$ stabilizer code is a quantum error correcting code that uses $n$ so-called data qubits to encode $k$ logical qubits with a code distance $d$. 
The code is specified (non-uniquely) by $n-k$ independent stabilizer generators that define the set of valid code-words.
Each stabilizer generator is an observable that is implemented by measuring an ancilla qubit (also called a \textit{measure} or \textit{syndrome} qubit) that has been coupled to a number of data qubits.
Assuming each stabilizer generator has its own measure qubit, the total number of physical qubits needed for an $[[n, k, d]]$ stabilizer code is $n_\text{phys}=n+(n-k)=2n-k$. 
A {\em stabilizer code cycle} or {\em syndrome extraction cycle} is a measurement of all $n-k$ stabilizers (typically implemented in parallel) which results in a collection of measurement outcomes called {\em syndromes} $(s\in\{+1,-1\}^{n-k})$.
When syndrome extraction cycles are repeated multiple times, changes in syndromes can be used to infer the most likely physical errors that have occurred in the code.

The surface code \cite{Fowler12} 
is a popular choice of stabilizer code because of its high error tolerance (between 0.1\% and 1\% for unbiased noise \cite{Stephens_14} and up to 43.7\% for biased noise \cite{Tuckett_18}) and because it can be implemented using only nearest-neighbor qubit interactions.
Crucially, (and unlike the majority of codes) it is also known how to perform universal quantum computation on surface code encoded qubits \cite{Litinski2019}.
In its original implementation, the surface code is a $[[d^2 + (d-1)^2, 1, d]]$ stabilizer code.
In this paper however, every mention of the surface code will refer to a similar (but slightly more efficient) variant called the \textit{rotated surface code} which has the parameters $[[d^2, 1, d]]$.


One of the necessary conditions for universality is the existence of an \textit{entangling gate}.
Several options exist for implementing two-qubit entangling gates between surface code qubits, though we limit our consideration to just two.
A \textit{transversal gate} is performed by coupling every physical qubit in one code to a counterpart in an adjacent code.
These transversal operations are challenging to implement in two-dimensional architectures due to the numerous \textit{non-local} interactions required \cite{Vasmer_19}.
An easier alternative is \textit{lattice surgery} which, unlike transversal gates, can be implemented with only nearest neighbor interactions \cite{Horsman_2012}.
In brief, lattice surgery is executed by performing a syndrome extraction over two code patches as if they were one elongated patch.
The disadvantage of lattice surgery is that, unlike a transversal gate, it requires measurements on a subset of the physical qubits that can introduce new, undetected errors into the ensemble.
To correct for this, it is necessary to perform at least $d$ rounds of lattice surgery to build confidence that the operation was done correctly.
This makes lattice surgery slow compared to the transversal gate but is still the favored option on account of its feasibility.

An important, though unrelated, fact is that transversal gates and lattice surgery can both be performed on surface codes that are not directly adjacent in space by using shared entanglement.
Specifically, maximally entangled qubit pairs can be used to teleport the required two-qubit gates in either case.
For surface codes of distance $d$, transversal gates require a total of $d^2$ pairs since each qubit in a $d \times d$ lattice must be matched up with its counterpart in the other lattice.
Coincidentally, lattice surgery also requires $d^2$ pairs since $d$ pairs are needed for each of the $d$ rounds that are necessary to account for measurement errors.
Although these two operations require the same total amount of entanglement, the \textit{rate} of required entanglement is much smaller for lattice surgery since only $d$ pairs are required at each time-step.

It was previously thought that the \textit{infidelities} of the entangled pairs used in lattice surgery must at least match the error threshold of the code, however recent work by Ramette et.\ al.\ indicates that pairs can tolerate an order of magnitude more error than expected \cite{ramette2023}.
The reason for this robustness stems from the fact that entangled pairs contribute errors along a single spatial dimension of the code during lattice surgery which limits the number of ways in which errors can form paths in the code.

\begin{figure}
    \includegraphics[scale=0.8]{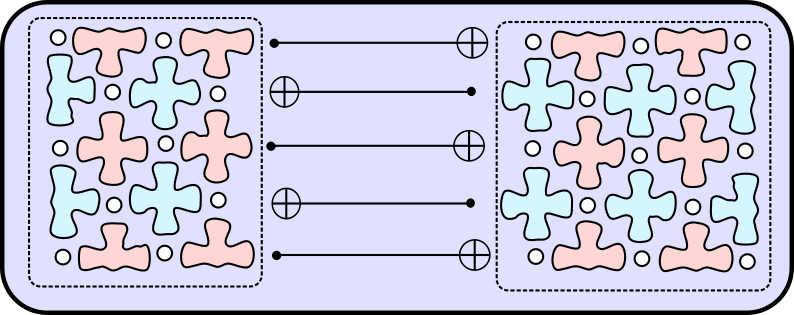}
    \caption{\label{fig:Surgery} 
    An illustration of the lattice surgery operation between two surface code encoded qubits.
    The white circles represent data qubits which work together to encode the logical qubits in either box.
    The floral tiles represent \textit{syndrome extraction circuits} which are executed in parallel to obtain information about errors that may have occurred on the data qubits.
    During a lattice surgery operation, two code patches undergo a syndrome extraction cycle together as if they were one elongated patch.
    For codes of distance $d$, lattice surgery requires $d$ entangled pairs along the seam to implement the $d$ CNOTs depicted.
    The effect of this operation is an $XX$ parity check between the two encoded qubits which, with single qubit gates, is sufficient for universal quantum computing.
    For an in-depth treatment of lattice surgery see \cite{Horsman_2012, chatterjee2024}}.
\end{figure}






\subsection{\label{sec:architectures}Trapped-ion architectures}

Monroe et.\ al.\ \cite{Monroe2014} were among the first to perform a systematic study of a modular trapped-ion architecture, and some of our our terminology follows from their work.
The modular trapped-ion computer is a collection of Elementary Logic Units (ELUs) which serve as local processors and or memory banks.
In the Monroe framework, each ELU is a linear Coulomb crystal containing an identical quantity of ions.
Some of these are \textit{communication ions} which are coupled to photonic interconnects and can be used to create entangled pairs between ELUs.
Still another fraction of the ions are may be used for sympathetic cooling, which is when one ion is brought in proximity to another to absorb some of its vibrational energy.

An attractive advantage of the linear crystal is that it is \textit{all-to-all} connected: No rearrangement is needed to implement a two-qubit gate between any pair of ions, though there's an effective limit to the crystal's length.
Monroe et.\ al.\ believed this limit to be around 100 ions which, ten years later, seems to be a well justified estimate since the longest linear crystal today is around 30 computational qubits \cite{Chen2023}.
Further corroborating evidence for this limit is discussed in greater detail by Murali et.\ al.\ \cite{Murali2020}.
Later research from the same Munroe group indicates that their initial 100 ion estimate might have been too optimistic \cite{Cetina_22}.
On the other hand, Ratcliffe et.\ al.\ \cite{Ratcliffe2018} propose a \textit{micro-trap} based architecture with all-to-all connectivity which they claim can be scaled beyond a linear trap.
This is accomplished with a \textit{non-adiabatic gate} which, at the cost of higher power, circumvents the need to use conventional side-band resolving gates which are slow and more susceptible to vibrational noise.
A two dimensional extention to the microtrap architecture has also been studied in some detail \cite{Mehdi_2020}.

In practice, we do not need to restrict ourselves by limiting the ELU to a single linear crystal. 
One possible workaround for further scaling is to have multiple crystals ``chained together" in the same ELU to form a \textit{segmented linear trap} \cite{Kaushal2020}.
The crystals in this chain can be separated, combined, and rotated to move a qubit in one crystal to any other crystal in the trap.
Though the segmented linear trap has a larger qubit capacity, it lacks the all-to-all connectivity of a single crystal -- meaning that some amount of physical routing is required for general computation.
On average, the number of swap operations needed to move quantum information from one end of the chain to the other scales linearly with the number of modules and the number of qubits per segment.  
Monroe et.\ al.\ believe more optimistically that a maximum of 1000 ions should be possible in an segmented linear trap \cite{Monroe2014}.
Recent work indicates that stabilizing ions with \textit{optical tweezers} could significantly improve the scalability of long ion-crystals \cite{schwerdt2024} \cite{Shen_2020}.

A still more general option for an ELU architecture is a Quantum Charged Coupling Device (QCCD) \cite{Kaushal2020} \cite{Pino2021}.
This is a two-dimensional configuration of ion chains which may be routed along a fixed network of corridors and junctions.
Theoretically, QCCD architectures are monolithic, well connected, and faster than segmented linear traps, though they are difficult to fabricate.
Presently, the largest QCCD is \textit{Quantiniuum's} 32 qubit \textit{race-track} computer,
which is a linear ion trap with periodic boundary conditions \cite{racetrack}.

Various theoretic proposals exist for scaling beyond this current best:
Malinowski et.\ al.\ report on a QCCD architecture called \textit{WISE} that addresses a crucial \textit{wiring problem} and is capable of supporting around 1000 ions \cite{malinowski_23}.
Valentini et.\ al.\ developed the so-called \textit{Quantum Spring Array} (QSA) where no inter-trap routing is required \cite{valentini2024}.
Sterk et.\ al.\ propose an architecture that specifically addresses the problem of power dissipation when scaling QCCDs \cite{sterk2024}.
Mehta et.\ al.\ suggest the use of planar-fabricated optics for further improvements \cite{Mehta_2020}.
For a survey on the technical challenges of scaling QCCDs, see Murali et.\ al.\ \cite{Murali2020}. 
True two and three dimensional ion crystals (also called Wigner crystals) may also be possible to engineer in the future on the scale of hundreds or perhaps even thousands of ions \cite{Wu_2021} \cite{Wang2015}.

\subsection{Trapped-ion surface code implementations}


Specialised architectures for \textit{quantum error correction} will likely be easier to realise in the near term 
since error correction is predictable, repetitive, and often nearest neighbor (as is the case for the surface code).
Recent experimental efforts to build trapped-ion surface codes are encouraging, though limited in scope.
Erhard et.\ al.\ for example used a ten ion quantum computer to perform quantum state teleportation between surface codes of distance two \cite{Erhard_2021} while Egan et.\ al.\
implemented the closely related Bacon-Shor code \cite{Egan_2021} based in part on theoretical results from \cite{Debroy_2020}.

There is also a growing body of theoretical work around this objective of building trapped-ion surface codes.
LeBlond et.\ al.\ presented software for compiling surface code operations to trapped-ion hardware \cite{LeBlond_23}.
Trout et.\ al.\ conducted extensive simulations of a distance 3 surface code implemented in a trapped-ion linear array \cite{Trout2018}, and reported a pseudothreshold of $3 \times 10^{-3}$ with syndrome cycle times ranging from around 3 to 8 milliseconds.
Similarly, Li et.\ al.\ studied a surface code modeled on a segmented linear trap \cite{Li2018}.
For segment lengths of around 15 qubits, they report an error tolerance of $0.12\%$ but say nothing about syndrome extraction times.
Lekitsch et.\ al.\ \cite{Lekitsch2017} present a proof-of-concept for a monolithic surface code architecture that closely resembles a QCCD.
A crucial difference however is that the Lekitsch architecture relies on \textit{global fields} for a majority of the operations instead of individual lasers, which they argue is more feasible for monolithic scaling since it circumvents the need to align many optical elements with high precision.
Their proposed cycle times are around $300 \mu s$.

Although the focus of this work is on the surface code, we note that there is a strong interest for the so-called \textit{color code} within the trapped-ion community \cite{Nigg_14, valentini2024, RyanAnderson_22}; 
This is a stabilizer code that is closely related to the surface code \cite{Kubica_2015}.

\section{Estimating surface code cycle times} \label{sec:estimating_cycle_times}


The rate we are able to do lattice surgery is upper bounded by the rate at which \textit{syndrome extraction cycles} can be performed.
Some care is therefore required to establish reasonable estimates for the attainable cycle times in trapped-ion systems.
Broadly speaking, there are three processes that need to be accounted for.
The first is the entangling operations that are required to couple the syndrome qubits with the appropriate data qubits.
Ions may or may not need to be routed for this depending on the underlying architecture.
The second process is the measurement of the syndrome qubits.
As we will soon see, this may require separating the ions a short distance from eachother in order to avoid measurement induced decoherence.
The third and final process is ion-cooling, which is necessary to ensure high fidelity two qubit operations.
Ion cooling may be a dedicated process during a surface-code cycle (after shuttling or measurement for example) or it may be continuous, as we will later discuss.
In brief, the time required for a surface code cycle will depend both on the underlying choice of architecture, but also on various technical factors such as the speeds of single and two-qubit gates, measurements, ion-shuttling and cooling.
We consider each of these factors in the following subsections and conclude our review by establishing three \textit{cycle time paradigms} we expect are feasible provided that specific technological milestones are met.

\subsection{Trapped-ion gates} \label{sec:gate_times}

The review of Bruzewicz et.\ al.\ presents a thorough comparison of single and two qubit trapped-ion gate times \cite{Bruzewicz19}.
Typical single qubit gates with fidelities greater than surface code threshold are reported between $2 \mu s$ and $12 \mu s$, though lower fidelity operations have been demonstrated on the order of nanoseconds.
Two qubit gates are generally slower and lower fidelity.
For the sake of argument, let us assume that we are willing to tolerate operational two-qubit error rates of up to 0.1\%, which sits comfortably below the surface code threshold.
Typical two-qubit gates with fidelities \textit{close} to  this rate are clocked between $1.6\mu s$ and.
It is not unreasonable to assume therefore that the time it takes to implement the gates in a surface code cycle is around $10 \mu s$.

\subsection{Trapped-ion measurements}

Single-qubit trapped-ion measurements are commonly implemented via \textit{state-dependent fluorescence}.
In this method, laser light is directed at an ion which exclusively couples the $|1\rangle$ state to a `cycling transition' that scatters numerous easily detectable photons.
Likewise, the absence of photons indicates a measurement of the $|0\rangle$ state \cite{Bruzewicz19}.
Although fluorescence measurements are relatively fast (on the order of $10\mu s$ \cite{Crain2019} \cite{Myerson2008}) and high fidelity, the scattered photons (both from the laser and from the irradiated ion) are likely to decohere nearby qubits that aren't also being measured.
This is a significant problem for error correcting circuits which all rely on mid-circuit measurements.
Broadly speaking, there are two complementary strategies for mitigating measurement induced decoherence.
The first is to incorporate techniques that suppress the decoherence, and the second is to move the ions some distance away to be measured safely.
Both of these strategies will be discussed in the following subsections.

\subsubsection{\label{sec:suppresing}Techniques for suppressing decoherence}

One way to limit measurement induced decoherence is to shorten the amount of time the qubit(s) are illuminated. 
This comes at the cost of measurement fidelity since there are fewer scattered photons to be detected.
Naturally, faster and higher fidelity measurements will be possible with improvements in the photon collection rate and photo-detector efficiency. 
See the introduction of Wolk et.\ al.\ for a brief summary of techniques used to improve sparse detection fidelities \cite{Wolk_2015}.

Another approach for protecting against decoherence is to use quantum logic spectroscopy \cite{Schmidt2005}.
This is when the information of one qubit is transferred onto an ion of a different species that when measured emits off-resonant photons which are unlikely to disturb the states of neighboring qubits.
An accidental benefit of this approach is that preexisting cooling ions may be used for this purpose.
The disadvantages of quantum logic spectroscopy are that it is more difficult to maintain coherent control of multiple ion-species simultaneously, and that it requires additional ions (at most double for a one-to-one pairing).

A promising alternative might be to use ions of the same species but have the data and measurement qubits encoded in different energy levels of the ion \cite{feng2023} \cite{Yang_2022}.
An alternative strategy would be to suppress the decoherence effects altogether so that additional measurement ions are not required.
Gaebler et.\ al.\ \cite{Gaebler2021} demonstrate a technique for reducing measurement cross-talk errors by an order of magnitude using tailored \textit{micromotion} which may reduce and potentially eliminate the need for logic spectroscopy or shuttling altogether.



\subsubsection{Ion-shuttling speeds}


Perhaps the most intuitive strategy for mitigating measurement induced decoherence is to move the ions a safe distance away before measuring them.
Ideally we'd like to complete this operation as fast as possible, but faster shuttling introduces more thermal noise which can have a detrimental effect on two-qubit gates in particular.
Broadly speaking, the infidelity of a M{\o}lmer-S{\o}rensen gate (See section \ref{sec:cooling}) applied between two qubits in an ion-chain is known to depend on both the temperature of the chain and its displacement in phase space. 
These effects have been fully characterised for two different error metrics \cite{Ruzic_2022}.
In the ion-shuttling literature, the amount of heat imparted in transport is commonly characterised in terms of how much the expected \textit{energy quanta} of a particular motional mode increases.
In the absence of noise, a linear crystal can withstand several quanta of phonons before there is an appreciable drop in the fidelity of a M{\o}lmer S{\o}rensen gate.
As noise is introduced however, this tolerance drops
\cite{Bentley_2020}. 
The fastest and quietest reported shuttling operation at the time of writing is also from Sterk et.\ al.\ who demonstrate a $210 \mu m$ one way ion transport in $6 \mu s$ with a maximum gain of $0.36 \pm 0.08$ quanta for an average speed of $35$ $\mu m$ $\mu s^{-1}$ \cite{Sterk2022}.
Slower, but more conservative routing was also demonstrated in $55 \mu s$ with a gain in 0.1 quanta \cite{Bowler2012}.

\subsubsection{Estimates for shuttling times}

How far do ions need to be separated for the effects of measurement induced decoherence to be considered negligible?
At the shorter end, Pino et.\ al.\ report a QCCD architecture where a shuttling distance of $110 \mu m$ resulted in cross-talk errors between $3.5 \times 10^{-3}$  and $1.5 \times 10^{-2}$ \cite{Pino2021}.
Similarly, Crain et.\ al.\ show that a separation of $370 \mu m$ results in cross-talk errors of $2 \times 10^{-5}$ \cite{Crain2019}.
If we assume a distance of $300 \mu m$ is tolerable, then with the shuttling speed reported by Sterk et.\ al.\, we can assume that a two-way shuttling time of around $10 \mu s$ is sufficient for eliminating decoherence effects.

\subsection{Cooling trapped ions} \label{sec:cooling}

All gates and operations of trapped ion quantum computers require low temperatures, but this is especially true of the two qubit gates since they depend on vibrational coupling which is highly sensitive to noise.
In the early days of trapped ion-quantum computing, two qubit gates required temperatures close to the ground state energy.
The breakthrough discovery of M{\o}lmer and S{\o}rrensen \cite{MolmerSorensen_1999} shifted this paradigm by introducing a gate that could operate at the \textit{Doppler temperature} -- the temperature regime attainable with \textit{Doppler cooling}.
In the following subsections, we present a brief review of the cooling techniques used to bring \textit{collections of ions} to Doppler and sub-Doppler temperatures -- endeavoring to report approximate cooling times wherever possible.
Although cooling single ions is considerably easier than cooling ion crystals (since there are fewer motional modes to be addressed \cite{Kang_2023}), it is unlikely that single-ion cooling will be a leading technology in the context of quantum computation.
This is because virtually all quantum architectures keep their computational ions organised in crystals.

At a high level, Doppler cooling works by shining a laser on an ion with a frequency just below what the ion will absorb.
When the ion moves towards the laser, the incoming light is blue-shifted with respect to the ion which causes it to absorb a photon and slow down.
Aside from Doppler cooling, other established laser based cooling techniques that operate under similar principles include resolved sideband cooling, Raman sideband cooling, and Electromagnetically Induced Transparency cooling (EIT) with the fastest of these being EIT.
Feng et.\ al.\ report cooling a 40 ion chain to a near ground state energy in under $300 \mu s$ \cite{Feng2020}
while Jordan et.\ al.\ reach similar temperatures for a 100 ion Penning trap within $200 \mu s$.
Some disadvantages of EIT are that it has a limited range of motional frequencies it can cool, and it is slow at cooling low frequency excitations \cite{Joshi_2020}.

Sympathetic cooling, where cold ions are brought into physical contact with computational ions, has been discussed in some detail in the previous sections. This is a well-established cooling method that remains a popular choice today – seeing use for example in the race-track architecture by Quantinuum \cite{racetrack}.
A disadvantage of sympathetic cooling however is that it is relatively slow compared to other techniques and requires the use of additional ions \cite{Lin_2013}.
An experimental demonstration showed that ion chains up to length 28 could be cooled to the global Doppler cooling limit using only two dedicated cooling ions of the same species \cite{Mao_2021}.
This paper reported \textit{relaxation times} (defined as the time required for the noise to settle within $5\%$ of noise of the initial state) between 10 and 100 $ms$.
Though these numbers are somewhat discouraging, one promising direction for further study is \textit{persistent cooling} where a number of sympathetic cooling ions are brought in perpetual contact with computational ions.
As the computation proceeds, the cooling ions are continuously chilled with Doppler cooling. 
This technique could lift the requirement for cooling processes that halt the computation.
Lin et.\ al.\ present an analysis of the dynamics of a linear array where a small subset of the ions are continuously cooled \cite{Lin2015}.
Additionally, a theoretic proposal for sympathetic cooling between one ion and a pre-cooled resource ion can be accomplished on the order of tens of microseconds, which may find some applicability in this context
\cite{Sgesser2020}.

\textit{Rapid exchange cooling} is a recently proposed alternative to sympathetic cooling that was suggested by Fallek et.\ al.\ in the context of QCCD \cite{Fallek2024}.
Here, coolant ions in a continuously chilled bank are shuttled to and from the computational ions.
The authors of this work perform a proof-of-concept experiment in which two calcium ions are cooled with a round-trip shuttling time of $107.3 \: \mu s$ which, in their words, is ``an order of
magnitude faster than typical sympathetic cooling durations.''

\subsection{Cycle time paradigms} \label{sec:cycle_time_paradigms}

At the beginning of this section (\ref{sec:estimating_cycle_times}), we mentioned that estimating attainable cycle times is crucial to our resource estimation task.
This is because the cycle time has a \textit{direct bearing} on the number of communication ions we require per ELU;
Faster cycles mean that more ions are required to collect the necessary entanglement in a shorter period.
In this section we synthesise our findings from the previous review by proposing three \textit{cycle time paradigms} $(1000\mu s, 100 \mu s, 10 \mu s)$ that we could reasonably expect to see achieved for trapped-ion surface codes given various technical assumptions.
For a short-hand summary of these paradigms see Table \ref{tab:paradigms}. The first and slowest cycle time we propose is around $1000 \mu s$.
This is several times faster than what was simulated by Trout et.\ al.\ \cite{Trout2018} and around three times slower than the architecture proposed by Lekitsch et.\ al.\ \cite{Lekitsch2017}.
Additionally, Egan et.\ al.\ present a distance three Bacon-Shor code where the $X$ stabilizers are measured in approximately $3ms$, which is within one order of magnitude of $1000\mu s$ \cite{Egan2021}.
This time scale permits some flexibility in routing and cooling options making it especially suitable for segmented-linear trap and QCCD architectures which require extensive use of both.
Here, we expect one or more stages of cooling per cycle and shuttling to avoid measurement induced decoherence.
The second time proposed is $100 \mu s$. 
This is a more optimistic regime, being several times faster than the EIT cooling times reported in Section \ref{sec:cooling}.
Because of this, we require that fewer than one dedicated round of cooling is made per clock cycle. 
There is considerably less flexibility permitted for routing or shuttling at this scale.
Dedicated zones for multi-qubit measurement will likely become significant time and heat savers as will sub-quanta shuttling.
Architectures that are likely to be viable in this paradigm include linear-traps, Wigner crystals, and QCCDs.
It is likely as well that at least one strategy for mitigating measurement induced decoherence will be employed (Section \ref{sec:suppresing}).
The final, and most optimistic regime is $10 \mu s$.
At this timescale, our clock cycle matches the two qubit gate times reported in Sec.\ \ref{sec:gate_times} meaning that no processes are allowed which interrupt syndrome extraction.
Persistent cooling is an absolute necessity here, as is an architecture that doesn't require any routing or shuttling.
Linear traps, Wigner crystals or micro-trap based architectures are the most likely candidates for this regime.

\begin{center}
\begin{table}
\begin{tabular}{||c | p{2.25in}||} 
 \hline
 Cycle time & System assumptions \\ [0.5ex] 
 \hline\hline
 $1000 \mu s$ &
 This cycle time is comparable to theoretical proposals in literature \cite{Trout2018} \cite{Lekitsch_17}, and is one order of magnitude from the syndrome extraction time of an experimentally realized distance 3 Bacon-Shor Code \cite{Egan2021}.
 (See section \ref{sec:cycle_time_paradigms} for more context.)
 \\
 \hline
 $100 \mu s$ & Less than one dedicated round of cooling per cycle. EIT cooling with sub-quanta shuttling required.
 Multi-qubit measurement stages recommended.
 Will likely incorporate at least one suppression technique discussed in Sec.\ \ref{sec:suppresing}.
 \\
 \hline
 $10 \mu s$  & 
 Virtually no shuttling allowed. 
 Persistent cooling that doesn't pause syndrome extraction cycle is essential. Multiple suppression techniques from Sec.\ \ref{sec:suppresing} will likely be used together.
 Purification circuits are low-depth with one round of measurement.
 \\ [1ex] 
 \hline
\end{tabular} \label{tab:paradigms}
\caption{A summary of three cycle time paradigms for trapped-ion surface codes and the various technological milestones required for each speed.}
\end{table}
\end{center}




\subsection{Entangling ion pairs}

Any modular architecture requires some means of communicating quantum information between the constituent ELUs.
A common approach is to establish \textit{maximally entangled pairs} of qubits between dedicated communication qubits to be used for quantum state teleportation.
First, an ion is pulsed to create an entangled pair between an internal state of the ion and an emitted photon.
Then, photons emitted from two different ELUs are routed together and fused with a polarization resolving Bell measurement that entangles the separated ions.
The maximum theoretical rate at which ion-ion entanglement can be established is fixed by a constant called the photon scattering rate, which is around 100 MHz according to Stephenson et.\ al.\ \cite{oxford_paper}.
The same authors note however that entanglement rates are far lower in practice (up to several kHz though with low fidelities \cite{Stockill_17}) primarily because of low photon collection efficiencies \cite{Monroe2014}.
The best ion coupling at the time of writing comes from the same paper cited previously, and reports 94\% fidelity pairs at an average rate of $182$ Hz and with a success probability of
$p_c = 2.18 \times 10^{-4}$ 
per attempt.

\section{Methodology}

\subsection{The lattice surgery cycle}

\begin{figure}
    \includegraphics[scale=0.8]{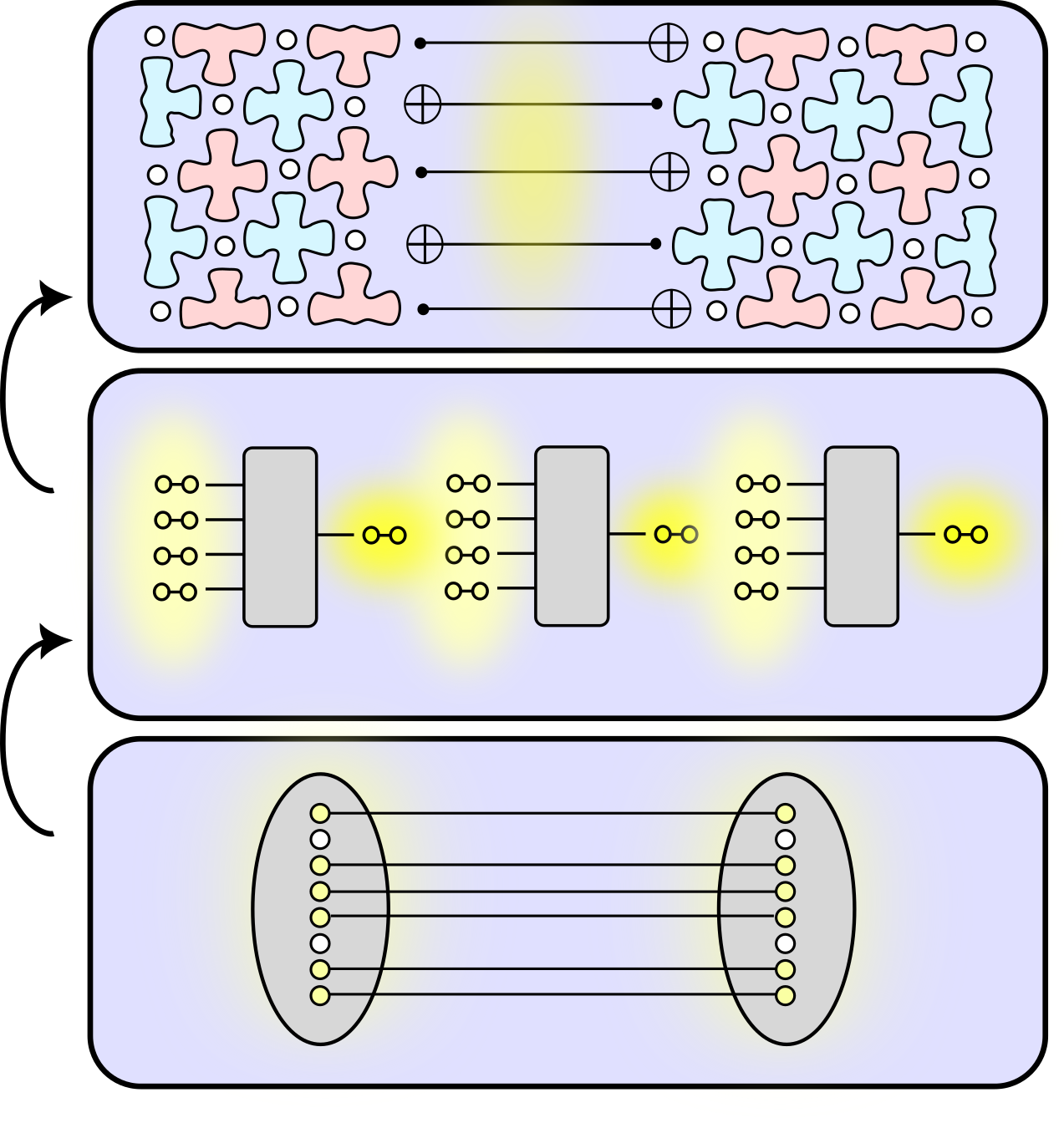}
    \caption{\label{fig:Protocol} 
    Three concurrent stages for a single lattice surgery operation that each must succeed within the cycle time $T$. (Bottom) Entanglement is established between pairs of communication ions in separate ELUs.
    (Middle) Distributed pairs in storage ions are refined via entanglement purification.
    (Top) Refined pairs are used to teleport the mediating gates needed for a lattice surgery between two surface code patches.
    }
\end{figure}

Given a surface code cycle time $T$, there are three steps that need to be completed within this $T$ for lattice surgery to be implemented.
The first is entanglement distribution, the second is entanglement \textit{purification} (see sec.\ \ref{sec:Purification}), and the third is the joint syndrome extraction.
All of these processes are illustrated in fig. \ref{fig:Protocol}.
A natural strategy is to complete these steps \textit{sequentially}.
This means we divide our time window $T$ into three parts which are then allocated to each process.
The disadvantage of this method however is that we cannot make use of the full time-window for any of the three steps.

The approach that we opt for instead is to implement these steps in \textit{parallel}.
In other words, the entanglement that we collect in one round is used for purification in the subsequent round which, in turn, is used for the surgery in the following round.
Although this gives us the leeway to implement each step within the timeframe $T$, there will necessarily be some \textit{dead time} when first starting the lattice surgery cycle as the initial entanglement is collected and processed.
We do not consider this dead-time in our analysis, but instead suppose that our lattice surgery ``engine'' is constantly running at a cycle time of $T$ with no starts or stops.

\begin{figure}
    \includegraphics[scale=0.50]{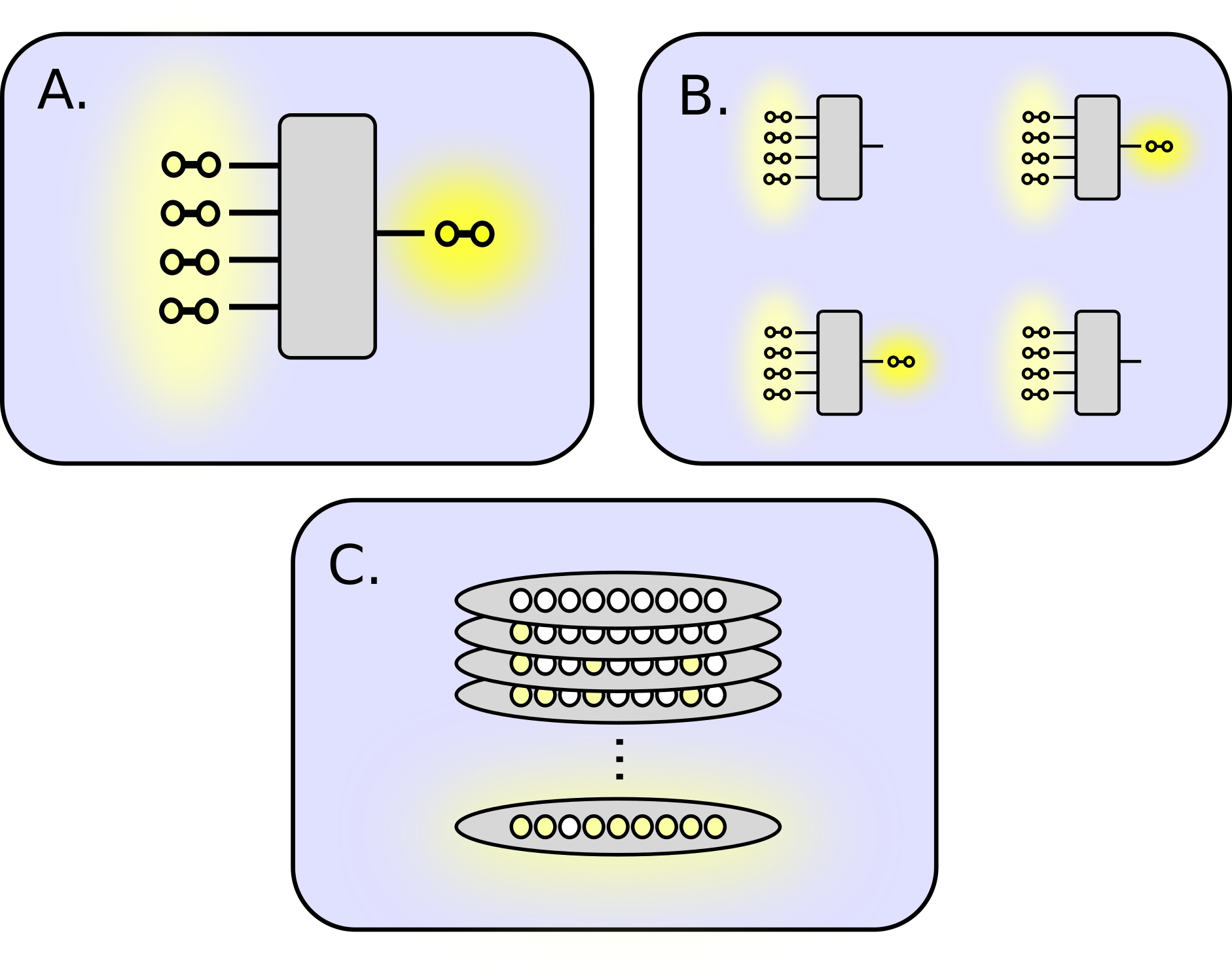}
    \caption{\label{fig:Heuristc} 
    Illustrations of the three conditions we require for our lattice surgery to be considered fault-tolerant: (A): We are given an $n\rightarrow1$ purification protocol where the output pair meets or exceeds fidelity $F_{\textrm{ideal}}$. 
    (B): We duplicate the circuit until the probability of getting at least one purified pair exceeds $P_{\textrm{pair}}$.
    (C): We have enough communication ions such that after after a collection time $T$, we are sure up to a confidence of $P_{\textrm{LS}}$ that we have enough entangled implement $D$ instances of the protocol described in (B).
    }
\end{figure}

\subsection{A heuristic for fault-tolerant lattice surgery} \label{sec:surgery_heuristic}

Our primary objective for this work is to estimate the number of trapped-ions needed to perform \textit{fault-tolerant} lattice surgery between surface codes at the rates specified in Table \ref{tab:paradigms}.
So far however, we have not addressed the question of what it means for lattice surgery to be fault-tolerant.
Here, we formalise a definition by introducing a heuristic consisting of three criteria that must all be met for lattice surgery to be considered fault-tolerant.
These conditions correspond to \textit{three subroutines} of lattice surgery, which are depicted in Fig.\ \ref{fig:Heuristc} (note that these correspond to the steps of fig \ref{fig:Protocol}).
Essentially, the point of this heuristic is to ensure that each of of these necessary subroutines succeeds with high probability.

The first criteria is the promise of a \textit{purification protocol} that takes $n$ Bell pairs of some initial fidelity $F_{in}$ and, with some probability $p$, returns one pair at or above the required fidelity threshold $F_{ideal}$.
The second condition is that each purification circuit is \textit{multiplexed} (run in parallel copies) until the probability of getting at least one pair from the lot exceeds $P_{\textrm{pair}}$.
The final condition is that we are able to collect enough entanglement within the cycle time $T$ to implement $d$ instances of the multiplexed purification protocol (one for each ``stitch" of the lattice surgery).
We stress that this collection \textit{must} take place within the time $T$ so that enough entanglement is acquired to be used for the next round of lattice surgery.
The probability that this collection succeeds must equal or exceed a threshold $P_{\textrm{LS}}$.


\subsubsection{Minimum number of communication ions required given cycle time}

Let $p$ be the success probability of a purification circuit.
The probability of obtaining at least one success out of $n$ trials is $1-(1-p)^n$.
Naturally the minimum number of purification circuits needed to produce at least one pair with a confidence of $P_{\textrm{pair}}$ is then

\begin{equation} \label{eq:multiplexing}
K \equiv \min_n  \bigg[ 1-(1-p)^n \geq P_{\textrm{pair}} \bigg]
\end{equation}

If the purification circuit takes $N_p$ raw pairs as input and returns one pair as output, the total number of raw pairs needed for the lattice surgery according to our heuristic is

\begin{equation} \label{eq:NLS}
    N_{\textrm{LS}} = dN_p K
\end{equation}

If $T$ is the surface code clock cycle (the time it takes to perform a round of syndrome extraction) and if $R$ is the rate at which entanglement can be attempted between pairs of ions, then we have $A = TR$ attempts to collect $N_{\textrm{LS}}$ pairs.
Suppose each ELU has $N_{\textrm{ions}} > N_{\textrm{LS}}$ communication ions.
During a collection attempt, each of the $v \leq N_{\textrm{ions}}$ vacant (unentangled) ions are pulsed and may become entangled with probability $p_e$.
Entangled ions are not pulsed in subsequent rounds, and we assume the entanglement does not degrade as it waits.
Our first objective is to determine the probability that $N_{\textrm{LS}}$ pairs can be collected in $A$ attempts.
The probability that a single ion pair is entangled after $A$ attempts is given by:

\begin{equation}
    P_{\textrm{onepair}} = 1 - (1-p_e)^A
\end{equation}

Let $X \sim \mathcal{B}(N_{\textrm{ions}}, P_{\textrm{onepair}})$ be the binomial random variable representing the number of ion pairs out of the initial $N_{\textrm{ions}}$ that are entangled after $A$ rounds. 
The minimum number of communication ions needed to collect at least $N_{\textrm{LS}}$ pairs in a code cycle with confidence $P_{\textrm{LS}}$ is then

\begin{equation} \label{eq:minIons}
    \min_{N_{\textrm{ions}}} \bigg [ P(X \geq N_{\textrm{LS}}) \geq  P_{\textrm{LS}} \bigg ] 
\end{equation}


\subsubsection{Maximum attainable rate given given number of communication ions}

Suppose now that $N_{\textrm{ions}}$ is fixed. 
Let $A_{\textrm{min}}$ be the minimum number of attempts needed to populate the $N_{\textrm{LS}}$ ions needed for lattice surgery with an overall confidence of $P_{\textrm{LS}}$.

\begin{equation} \label{eq:maxRate}
    A_{\textrm{min}} = 
    \min_{A} \bigg [ P(X \geq dN_p K) \geq  P_{LS} \bigg ] 
\end{equation}

The maximum attainable rate for our fault-tolerant lattice surgery given $N_{\textrm{ions}}$ is then just $A_{min} / R$

\subsection{Device parameters and assumptions} \label{sec:DeviceAssumptions}

From Stephenson et.\ al.\ \cite{oxford_paper}, we assume that we can pulse ions at a rate of $1 \textrm{MHz}$ where each pulse has a $2.18 \times 10^{-4}$ chance of producing an entangled ion-ion pair of fidelity $0.94$.
We assume that our surface codes have an operational error rate of $0.1\%$ which, from the results of Ramette et.\ al.\ \cite{ramette2023}, means we can tolerate Bell pairs with an infidelity of $0.01$.
We assume that the routing and circuits within the purification stage take a negligible amount of time compared with the entanglement collection.
Single and two qubit gate error are approximated as single and two-qubit depolarizing channels that occur with probabilities $1 \times 10^{-5}$ and $5 \times 10^{-5}$ respectively.
Measurement errors are taken as bitflip channels that occur with probability $1 \times 10^{-5}$.
Although ion-trapping lifetimes are extremely good (hours, and even months in extreme cases), they are not indefinite. 
We do not consider ion loss or replacement in our resource estimation.
Neither do we consider leakage errors that are known to accumulate with consecutive surface code cycles \cite{Brown2019}.
Though a linear crystal architecture has all-to-all connectivity in theory, it may not be possible in practice to perform arbitrary simultaneous two qubit gates as we have assumed.
Nevertheless, we note promising results from the current state-of-the-art \cite{Grzesiak2020}.

\subsection{Optimizing entanglement purification with device level noise} \label{sec:Purification}

\begin{figure*}
	\centering
	\includegraphics[width=\linewidth]{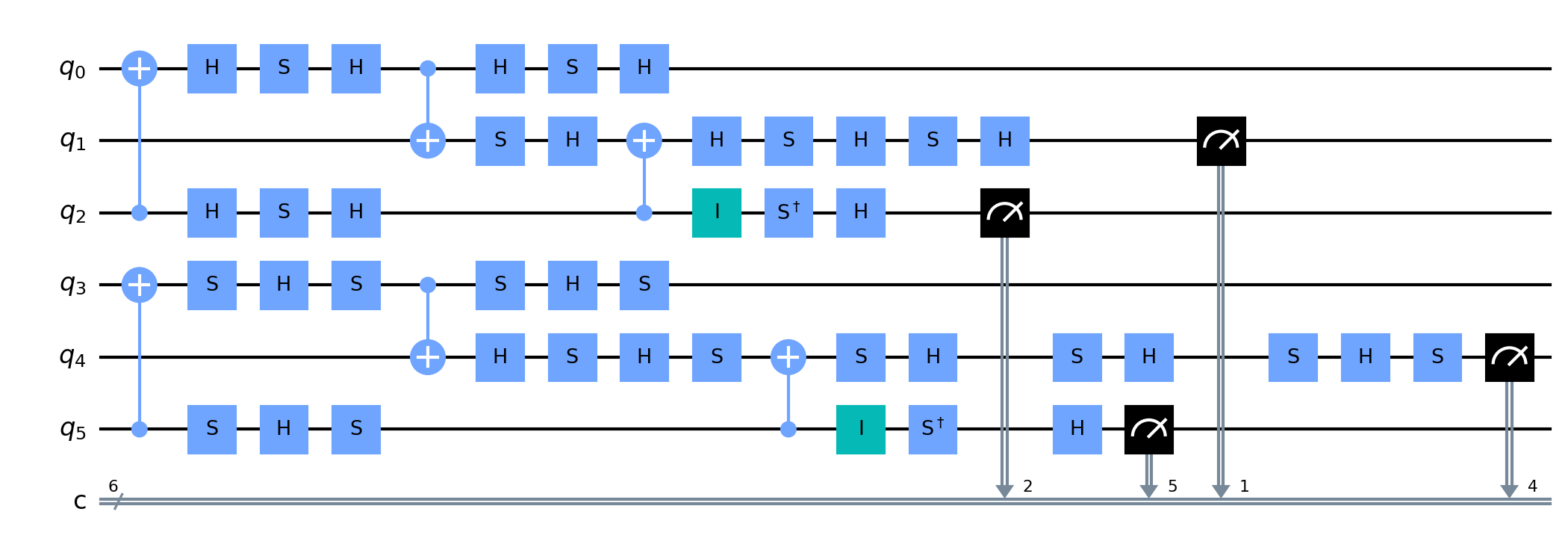} \\ 
\caption{A high yield purification circuit discovered by the genetic algorithm from Addala et.\ al.\ \cite{Addala_2023}.
 This circuit takes three partially entangled pairs as input $\{(q_0, q_3), (q_1, q_4), (q_2, q_5)\}$ and non-deterministically returns a higher fidelity pair at $(q_0, q_3)$.
 The protocol succeeds if $c_1 = c_4$ and $c_2 \neq c_5$ (in other words, the measurement outcome of qubit $q_1$ is coincident with the measurement of $q_4$ and the measurement of $q_2$ is \textit{anti}-coincident with $q_5$). 
 When this circuit is run with three copies of the \textit{Stephenson pair} (see Appendix) as input together with the noise parameters described in sec. \ref{sec:DeviceAssumptions}, the purification produces a fidelity $F=0.9904$ pair with probability $p = 0.819$}
	\label{fig:high_yield_purification}
\end{figure*}

Our need for fault-tolerant lattice surgery highlights the importance of a high yield pair purification protocol.
Though all such protocols theoretically asymptote to unit fidelity, the practical reality is that device level-noise imposes a cap on the pair fidelities that are attainable with purification.
It is essential therefore to find a purification protocol that is able to reach our target fidelity of $F_{\textrm{ideal}} = 0.99$ despite circuit-level noise.
To this end, we decided to search for high-performing purification protocols using recently developed numerical methods.
Goodenough et.\ al.\ proposed an exhaustive search over purification protocols by mapping the problem to an enumeration over so called \textit{graph codes} \cite{goodenough2023}, while Addala et.\ al.\ \cite{Addala_2023}, built on earlier work from Krastanov et.\ al.\ \cite{Krastanov2019optimized} to refine a genetic algorithm for finding purification protocols.
We opted to use the genetic optimization over the enumeration because of its ease of use and because the attainable fidelities reported by Addala et.\ al.\ were comparable to those reported by  Goodenough.
We used this genetic algorithm to identify 
several hundred potentially suitable purification circuits.
From this initial pool of candidates, 
we simulated each circuit using the noise parameters detailed in \ref{sec:DeviceAssumptions}.
For added realism, we modeled our initial $F = 0.94$ entangled pairs after the density matrix of an experimentally realised ion-ion pair reported in the supplementary material of Stephenson et.\ al.\ \cite{oxford_paper} (See appendix).
The highest yield purification circuit we discovered with this genetic algorithm is presented in fig.\ \ref{fig:high_yield_purification}.
This protocol takes three \textit{Stephenson pairs} as input and produces one output pair with a fidelity of $F = 0.9904$ with an overall success probability of $0.819$.
A full discussion of our methodology and findings is presented in the appendix of this paper.


\section{Results}

\begin{figure}
    \includegraphics[scale=0.55]{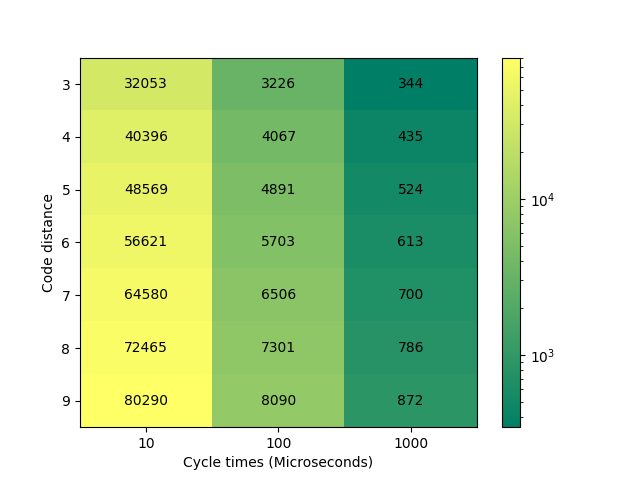}
    \caption{\label{fig:MinIons} 
    The minimum number of communication ions that are required to perform fault-tolerant lattice surgery (see section \ref{sec:surgery_heuristic}) for a range of code distances and surface code cycle times.
    }
\end{figure}

\begin{figure}
    \includegraphics[scale=0.55]{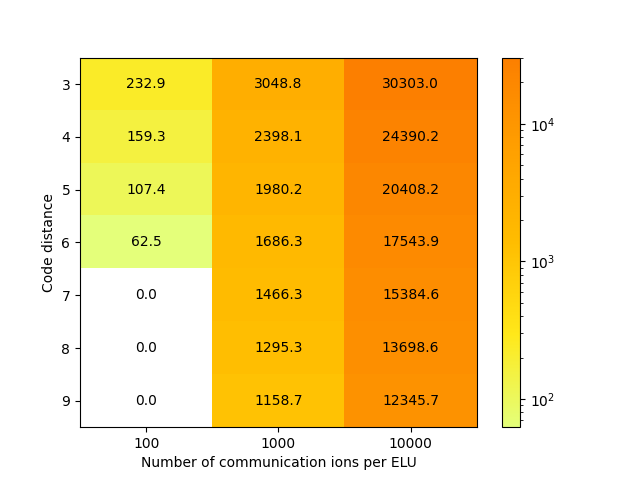}
    \caption{\label{fig:LSRates} 
    Average fault-tolerant lattice surgery rates (in Hertz) for different code distances and numbers of communication ions. The squares marked with ``0.0" signify that fault tolerant lattice surgery is not possible at the specified parameters.
    }
\end{figure}

Our numerical results are presented as tables in Fig.\ \ref{fig:MinIons} and Fig.\ \ref{fig:LSRates} respectively.
In Fig.\ \ref{fig:MinIons} we used Eq.\ \ref{eq:minIons} to determine the minimum number of communication ions needed to collect sufficiently many entangled pairs for fault-tolerant lattice surgery for a given cycle time $T$.
We considered three cycle times of $10 \mu s, 100 \mu s,$ and $1000 \mu s$ according to the technological paradigms we discussed in section \ref{sec:estimating_cycle_times} over a small range of code distances.
Our calculations indicate that the number of communication ions required is approximately linear with respect to both the code distance and the cycle time within our selected ranges.
As the cycle time decreases in orders of magnitude, we find straightforwardly that the number of communication ions increases in orders of magnitude.
If we assume that a given ELU may contain around 1000 ions at most, we find that a $d = 9$ code is theoretically supported at a clock cycle $1000 \mu s$, while cycle times considerably faster than this appear out of reach.

In fig \ref{fig:LSRates}, we used Eq.\ \ref{eq:maxRate} to calculate the maximum lattice surgery rates that are theoretically possible according to our fault-tolerant heuristic given various numbers of communication ions and various code distances.
The whited out squares in the 100 ion column from $d = 7$ onward indicate that fault-tolerant lattice surgery is not possible at these distances, since the number of required pairs exceeds the number of communication ions available.
Similar to what we observed in the previous table, we find that there's a tenfold difference between the 1,000 and 10,000 ion columns, though we note a slight deviation from this trend at the 100 ion column.
This behavior occurs because the number of communication ions is close to the number of required pairs.
As the required number of pairs approaches the number of available ions, we expect an exponential increase in the number of attempts needed to collect the entanglement since this collection is done without replacement.
Our results indicate that with 100 communication ions, we could expect to support a distance 5 or 6 code at a maximum rate of around $100$Hz.
For 1000 and 10000 communication ions we find that larger code distances are possible with rates at around $1$KHz and $10$KHz respectively.

\begin{figure*}
    \centering
    \includegraphics[scale=0.85]{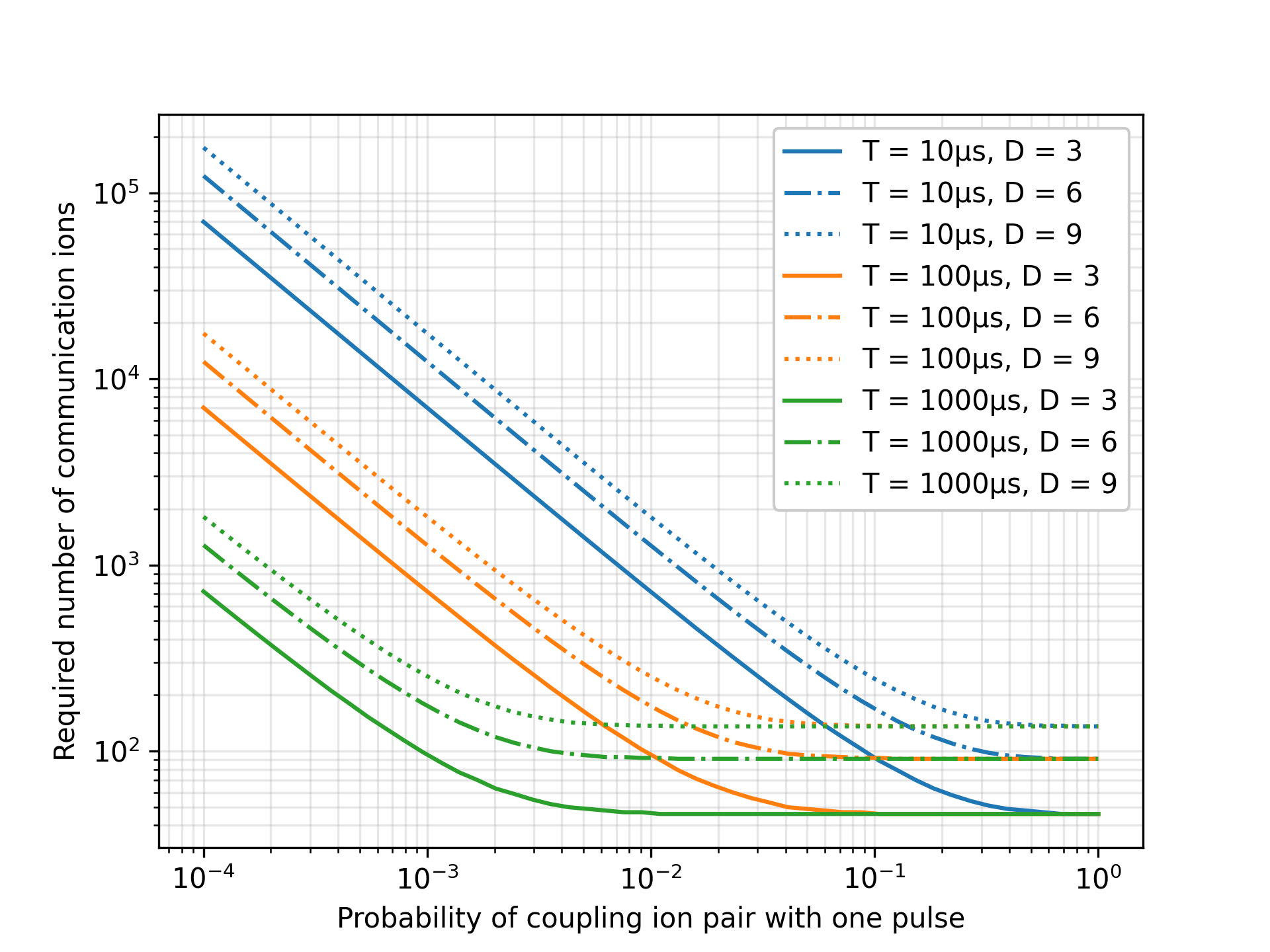}
    \caption{
    A selection of line plots that relate the minimum number of communication ions needed per trap 
    with the probability of entangling two inter-trap ions with a single \textit{`pulse'}.
    The three line colors represent the three surface code cycle times proposed in section \ref{sec:cycle_time_paradigms} while the three linestyles represent various surface code distances.
    On the left, the data points begin at probability $p_c = 10^{-4}$, which is comparable to the best known coupling odds of $p_c = 2.18 \times 10^{-4}$ reported by Stephenson et.\ al. \cite{oxford_paper}.
    Fewer communication ions are needed as the success probability increases since it becomes easier to establish entanglement links between traps.
    At high probabilities, the lines converge to the same three values for the respective distances;
    These values represent the minimum number of unrefined pairs that are required for entanglement distillation to yield enough $F = 0.99$ links for lattice surgery.
    For $d = 3, \: 6$ and $9$, these values are $46, \: 91,$ and $136$ respectively.
    }
    \label{fig:ionvscouplingrate}
\end{figure*}


Our results indicate long term concerns for scalability due to the large number of physical resources required.
If $10 \mu s$ is about the fastest clock cycle we can hope for, our findings in figure \ref{fig:MinIons} indicate that we would need upwards of 40,000 communication ions per ELU even for modestly sized surface codes.
The prohibitive cost of this indicates an urgent need for improved optical coupling;
As it stands, the low probability of entangling ions in separate ELUs ($p_c = 2.18 \times 10^{-4}$  \cite{oxford_paper}) is the leading cause for this inflated resource overhead.
A natural question then (and the focus of this section) is how the number of communication ions scales with \textit{improvements} in the coupling rates.
Our results are presented in figure \ref{fig:ionvscouplingrate}.
Here, we have again used equation \ref{eq:minIons} to calculate the minimum number of ions needed for lattice surgery between a selection of different surface codes
while varying the probability $p_c$ of establishing entanglement between a given pair of ions.
What we see is that, irrespective of the distances and cycle times we consider, the number of communication ions first decreases as a power law with increasing $p_c$ and eventually tapers off to a fixed value that depends on the code distance and the purification protocol.
These plateau values are equal to $N_{LS} = d N_p K$, which is the number of unpurified pairs that are required between ion-traps for lattice surgery to succeed (see eq.\ \ref{eq:NLS}).
We can easily demonstrate this by examining the limiting case behavior when $p_c = 1$. In this scenario, all ions are guaranteed to be entangled after a single round of attempts. 
Consequently, we only require $N_{LS}$ many ions to ensure that at least $N_{LS}$ many raw pairs are established between traps.

For the sake of transparency, we note that the number of communication ions reported at the plateaus of fig.\ \ref{fig:ionvscouplingrate} are $46, \: 91,$ and $136$ for distances $d = 3, 6$ and $9$ respectively. We point this out because for $K = 5$ and $N_p = 3$, these values are equal to $N_{LS}+1$.
This extra $+1$ comes from a small programming oversight where eq.\ \ref{eq:minIons} was calculated using a strict inequality as opposed to a weak inequality.

The data of figure \ref{fig:ionvscouplingrate} additionally allows us to pin-point the ion coupling probabilities at which we reach these plateaus. \st{and consequently have no need to improve further}
For a cycle time of $1000 \mu s$, this occours at around $p_c \approx 10^{-2}$, while for $100 \mu s$ and $10 \mu s$ we find the convergence points at $p_c \approx 0.1$ and $p_c \approx 0.5$ respectively.

Let us suppose for the sake of argument that we are allowed 200 communication ions per trap; 
This is high, but (unlike our data in figure \ref{fig:MinIons}) is not outside the realm of possibility.
From figure \ref{fig:ionvscouplingrate}, the ion coupling probabilities that are required to perform lattice surgery for a distance 9 surface code at cycle times of $1000 \mu s, 100 \mu s$ and $10 \mu s$ respectively are $p_c \approx 1.5 \times 10^{-3}$, $p_c \approx 1.5 \times 10^{-2}$ and $p_c \approx 1.5 \times 10^{-1}$.
These data suggest that ion-coupling rates need to improve by \textit{one or several} orders of magnitude depending on the cycle time one wishes to operate at.
One way to help meet this demand is to improve the efficiency of photon collection.
Carter et.\ al.\ report collection efficiencies of $10\%$ which is roughly an order of magnitude above what was previously possible \cite{Carter2023}.
Based on this result, it seems that improving entanglement rates by an order of magnitude is feasible target.
Whether further improvements are possible however is unclear to us at present.
Alternative methods for transporting entanglement via \textit{shuttling} (therefore bypassing the need for improved coupling) are conceivable, yet speculative.
Entanglement distribution using neutral atoms to mediate interactions has been discussed in the context of quantum networking \cite{Hannegan2021}.

Entanglement purification is another target for improvement.
Our $3 \rightarrow 1$ protocol has a $81.9\%$ success rate, which for a confidence threshold of $P_{\textrm{Pair}} = 0.999$ means that we require $K = 5$ purification circuits (Eq.\ \ref{eq:multiplexing}) per `stitch' in the lattice surgery.
This is effectively a $15 \rightarrow 1$ purification circuit, which seems somewhat wasteful.
Given that this protocol is likely close to optimal for our initial state, the most likely strategy for reducing these overheads is to improve the fidelity at which pairs are distributed.
Ideally, purification is eliminated altogether by delivering pairs a fidelity of $F = 0.99$ or higher.



\section{Conclusion}

In this work we estimated the number of communication ions needed to perform lattice surgery between two trapped-ion surface code qubits.
To this end, we developed three paradigms for syndrome extraction cycle times that are predicated on various technological milestones and presented a heuristic 
that establishes what it means for a lattice surgery operation to be fault tolerant.
With current inter-trap coupling rates, we find that hundreds, thousands and tens of thousands of communications ions are required for fault tolerant lattice surgery at cycle times of $1000 \mu s, 100 \mu s$ and $100 \mu s$ respectively.
The primary factor contributing to these prohibitive overheads is poor ion-coupling rates.
Our results indicate the need to improve the coupling probability $p_c$ by at least an order of magnitude for lattice surgery to be possible with only a couple hundred resource ions.


\section{Acknowledgement and funding}


We thank Ilia Khait, David Elkouss, Stefan Krastanov, Vaishnavi Addala, and Kenneth Goodenough for helpful discussions. 
Additional thanks are extended to the Centre for Quantum Computing and Communication Technology (CQC2T).
The views, opinions, and/or findings expressed are those of the author(s) and should not be interpreted as representing the official views or policies of the Department of Defense or the U.S. Government. This research was developed with funding from the Defense Advanced Research Projects Agency [under the Quantum Benchmarking (QB) program under Awards No. HR00112230007 and No. HR001121S0026 contracts].

\bibliography{paperbib.bib}

\clearpage
\appendix*

\section{Noisy entanglement distillation} \label{sec:Apdx_search}

Implementations $\tilde{\mathcal{D}}$ of any entanglement distillation protocol $\mathcal{D}$ are generally subjected to noise in the sense that $0\neq\lVert\tilde{\mathcal{D}}-\mathcal{D}\rVert\leq\epsilon$ for small $\epsilon$. Here we benchmark the performance of entanglement distillation protocols obtained with genetic algorithm~\cite{Addala_2023} subjected to noise in ion trap systems. The genetic algorithm takes a Bell-diagonal state
\begin{align*}
    F\phi_+ +
    (1-F)(p_x\psi_+ + p_z\phi_- + p_y\psi_-)
\end{align*}
as input and searches for optimal $n \rightarrow k$ purification protocols for that state by iterating an initial randomly generated population of circuits (describing entanglement distillation protocols) over a number of generations.
The fitness of the individuals in the population is evaluated with respect to one of several possible objective functions.
We optimized with respect to the ``average marginal fidelity," which is the average fidelity of each output pair traced out from the final ensemble.
This is determined analytically and may optionally account for single and two qubit depolarizing gate noise along with measurement errors.
Each output circuit is returned with its average marginal fidelity and overall success probability.

Although the Bell-diagonal states may at first appear to be a somewhat contrived category of entangled pairs, it turns out that all two-qubit mixed states can be deterministically \textit{twirled} into Bell-diagonal pairs of the same fidelity using local operations and classical communications.
This may however cost a small amount of the \textit{distillable entanglement} depending on the input state, though quantifying the amount of entanglement lost and developing recovery techniques appear to be open research directions.
For this reason, and because twirling introduces a small amount of noise, we developed our search methodology around the objective of finding purification protocols that could work without twirling.

\begin{figure*}
    \includegraphics[scale=0.70]{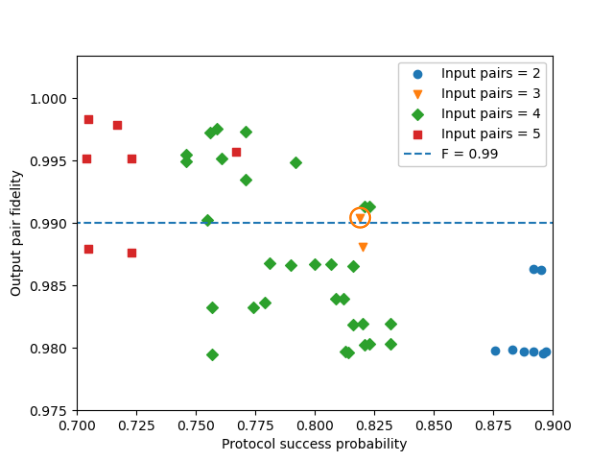}
    \caption{\label{fig:PurificationStatistics} 
    A scatter plot of the success probabilities and output pair fidelities
    of a high performing subset of $n \rightarrow 1$ purification protocols that were first identified with the genetic algorithm~\cite{Addala_2023} then simulated under circuit level noise with $F=0.94$ Stephenson pairs (See eq.\ \ref{eq:rotated_stephenson}) as inputs.
    The highest yield purification protocol that exceeds the required fidelity threshold $F_{\textrm{ideal}} = 0.99$ is circled in orange.
    This is a $3 \rightarrow 1$ purification protocol whose circuit is presented in fig.\ \ref{fig:high_yield_purification}.
    }
\end{figure*}

Our strategy therefore was to perform an initial search for promising looking protocols using $F = 0.94$ Bell-diagonal pairs with $p_x = p_y = p_z = 1/3$ under the parameter values $n = (3,4,5)$ and $k = 1$.
Our decision to limit our search to $k = 1$ was motivated by our analysis indicating that the $k > 1$ protocols produced output pairs with some amount of \textit{mutual entanglement} between them.
This is a significant issue for lattice surgery, since it is assumed that each input pair is independent and required us to restrict ourselves to the $k = 1$ case.
Simulating beyond $n = 5$ proved to be unnecessary since the average success rate of the protocols can be seen to decrease with increasing $n$ in Fig.\ \ref{fig:PurificationStatistics}.
For an $(n > 4) \rightarrow 1$ protocol to have a \textit{higher yield} than the $3 \rightarrow 1$ we identified, it would be necessary for the larger purification circuit to have a significantly higher success probability.

Each simulation was performed with a population of 100 circuits evolved for 150 generations.
We selected the top performing circuits from each simulation 
and benchmarked their performance under the same circuit level noise when simulated with $F = 0.94$ \textit{Stephenson pairs} (See sec.\ \ref{sec:Purification}) as input.
Our numerical results are presented in Fig.\ \ref{fig:PurificationStatistics}.
The broad trend indicates that as the number of pairs increases, the success probability of the protocol decreases while the average success probability increases.
None of the protocols we identified were able to exceed the required fidelity threshold of $F_\text{req} = 0.999$.

\section{The Stephenson pair} \label{app:stephenson_pair}

In section \ref{sec:Purification}, we alluded to an experimentally realised inter-trap ion pair reported by Stephenson et.\ al.\ in the supplementary material of their main paper \cite{oxford_paper}.
Strictly speaking, there are \textit{four pairs} reported which correspond to four possible interferometer detection events.
Since these state are all effectively equivalent under local operations and classical communications, we arbitrarily chose to consider the state presented in equation \ref{eq:unrotated_stephenson}.
When expressed in the Bell basis, the predominant term of this state is seen to be $|\phi^-\rangle \langle \psi^-|$ up to some phase. As a matter of convenience, it is preferable for this state to be predominantly entangled with respect to $|\phi^+\rangle \langle \phi^+|$.
We therefore apply the rotation $(I \otimes XZ) \rho (I \otimes XZ)$ to obtain $\rho'$ in equation \ref{eq:rotated_stephenson}. This is the \textit{Stephenson pair} that we refer to in the main body of our paper.

\begin{figure*}
\begin{equation} \label{eq:unrotated_stephenson}
\rho = 
\left(
\begin{array}{cccc}
 0.01 & -0.00487616+0.00349614 i & 0.0135924\, +0.00634402 i & 0.00374015\, -0.00331833 i \\
 -0.00487616-0.00349614 i & 0.569 & 0.0542638\, +0.440672 i & -0.012985-0.0292471 i \\
 0.0135924\, -0.00634402 i & 0.0542638\, -0.440672 i & 0.416 & -0.0225074-0.00473484 i \\
 0.00374015\, +0.00331833 i & -0.012985+0.0292471 i & -0.0225074+0.00473484 i & 0.005 \\
\end{array}
\right)
\end{equation}
\caption{The ``unrotated" Stephenson state, taken from Fig.\ S4(i) of the supplementary material of \cite{oxford_paper}.}
\end{figure*}

\begin{figure*}
\begin{equation} \label{eq:rotated_stephenson}
\rho' = 
\left(
\begin{array}{cccc}
 0.569\, +0. i & -0.00487616-0.00349614 i & -0.0292471+0.012985 i & 0.440672\, -0.0542638 i \\
 -0.00487616+0.00349614 i & 0.01\, +0. i & -0.00331833-0.00374015 i & 0.00634402\, -0.0135924 i \\
 -0.0292471-0.012985 i & -0.00331833+0.00374015 i & 0.005\, +0. i & -0.0225074+0.00473484 i \\
 0.440672\, +0.0542638 i & 0.00634402\, +0.0135924 i & -0.0225074-0.00473484 i & 0.416\, +0. i \\
\end{array}
\right)
\end{equation}
\caption{A rotated version of Eq.\ \ref{eq:unrotated_stephenson} obtained with the transformation $\rho' = (I \otimes XZ) \; \rho \; (I \otimes XZ)$}
\end{figure*}

\section{Tables of constants}

In this section we present two tables that detail the most important free parameters and physical constants used throughout the paper.
Table \ref{tab:FreeParams} is a summary of the free parameters used and table \ref{tab:Constants} is a summary of the constants.

\begin{table*}
  \begin{tabular}{ ||c|c|| } 
    \hline
    Parameter & Definition 
    \\ 
    \hline \hline
    
    $N_{\textrm{ions}}$ & The number of communication ions in an ELU \\
    \hline
    
    $d$ & Code distance \\
    \hline
    
    $T$ & Syndrome extraction cycle time  \\  
    \hline
    
  \end{tabular}
  \caption{A summary of the free parameters considered in our analysis.}
  \label{tab:FreeParams}
\end{table*}

\begin{table*}
  \centering
  \begin{tabular}{ ||c|p{2.5in}||c|p{2.5in}|| } 
    \hline
    Parameter & Definition & Value & Justification \\ 
    \hline \hline

    $p$ & Purification protocol success probability & 0.819 & Simulated numerically under circuit level noise \\
    \hline

    $R$ & Pulse rate & $1$ MHz & Within the magnitude of what is physically possible \cite{oxford_paper} \\
    \hline
    
    $p_c$ & The probability of entangling two ions with one pulse-attempt & $2.18 \times 10^{-4}$ & State of the art: \cite{oxford_paper} \\
    \hline

    $F_{\textrm{ideal}}$ & The fidelity required for pairs used in lattice surgery & 0.99 &  
    Originally $0.999$, but improved thanks to \cite{ramette2023} \\
    \hline

    $N_p$ & The number of pairs required for the purification circuit & 3 & Fig.\ \ref{fig:high_yield_purification}  \\
    \hline

    $P_{\textrm{pair}}$ & The required confidence for multiplexed purification circuits to produce at least one pair & 0.999 & Comfortably below the surface code threshold \\
    \hline

    $K$ & The required number of purification circuits needed to meet multiplexing confidence & 5 & Substituting appropriate values into eq.\  \ref{eq:multiplexing} \\
    \hline

    $P_{\textrm{LS}}$ & The required confidence for collecting sufficient entanglement within a given clock cycle & 0.999 &
    For a surface code of distance $d \leq 9$ (the largest distance we consider in our analysis), A $P_{LS}$ of $0.999$ is sufficient to guarantee that a logical CNOT between surface codes can be implemented with success probability greater than $99 \%$.
    This is seen by observing that $(0.999)^9 > 0.99$.
    \\
    \hline
    
  \end{tabular}
  \caption{A catalogue of important numerical constants used throughout this paper with justifications.}
  \label{tab:Constants}
\end{table*}

\end{document}